\documentclass[%
 reprint,
 amsmath,amssymb,
 aps,
]{revtex4-2}

\usepackage{graphicx}
\usepackage{dcolumn}
\usepackage{bm}
\usepackage[version=4]{mhchem}

\begin{document}

\preprint{APS/123-QED}

\title{Abnormal improved dielectric in La doped relaxor ferroelectric  PNN-PHT
}

\author{Shiyi Zhang}
\author{Dongyan Zhang}%
 \email{zhangdongyan@xidian.edu.cn}
 \author{Zhimin Li}
 \email{zmli@mail.xidian.edu.cn}
\affiliation{%
 School of Advanced Materials and Nanotechnology, Xidian University, Xi'an 710126, P. R. China
}%

\author{Pangpang Wang}
\affiliation{
 Nanomaterials Group, Institute of System, Information Technologies and Nanotechnologies (ISIT), Fukuoka Industry-Academia Symphonicity (FiaS), 4-1 Kyudaishinmachi, Nishi-ku, Fukuoka 819-0388, Japan
}
\author{Ri-ichi Murakami}
\affiliation{%
 School of Mechanical Engineering, Chengdu University, Chengdu 610106, P. R. China
}%

\date{\today}

\begin{abstract}
Relaxor ferroelectrics are complex materials with distinct properties different from classic ferroelectrics. Probing frequency depended dielectric susceptibility is its trait. Other empirical laws build on classic ferroelectrics are still applicable to relaxor ferroelectrics. Here we report an abnormal improved dielectric in La doped relaxor ferroelectric  0.4PNN-0.6PHT, which deviated from the consistent variation relationship with piezoelectric coefficient. A mesoscale mechanism is proposed to reveal the origin of the improved dielectric response in 0.4PNN-0.6PHT, where the polar nanoregions alignment can facilitate polarization response to external field. This mechanism emphasizes the critical role of polarization fluctuation on the macroscopic properties of 0.4PNN-0.6PHT.

\end{abstract}

\maketitle

\section{\label{sec:Intro}Introduction}
Relaxor ferroelectrics are complex materials with distinct properties different classic ferroelecctrics, which was named due to the diffuse phase transition arsing from the compositionally  disorder \cite{BOKOV2012JAD, ZHANG2021109447, ZhangPRB2021}. Various compounds were found to posses the feature of diffuse phase transition, such as $Pb(Mg_{1/3}Nb_{2/3})O_3-PbTiO_3$ (PMN-PT), $Pb(Ni_{1/3}Nb_{2/3})O_3-Pb(Hf, Ti)O_3$ (PNN-PHT), and etc \cite{LiFeiNatMat2018,LiFeiScience2019,YangxiACSInterface2021}. Relaxor ferroelectrics have  been widely investigated  due to their giant piezoelectricity overtaking classic ferroelectrics \cite{LiFeiNatMat2018,LiFeiScience2019}. Although a lot of distinctive properties were found in relaxor ferroelectrics, including frequency dispersion of permittivity \cite{ChenJAP2019}, deviation from Curie–Weiss law in the vicinity of $T_m$ \cite{BOBIC2018233}, inhomogeneous cation arrangement \cite{ZhangSciRep2020}, polar nanoregions \cite{EremenkoNatCommun2019}, and etc.,  there are still some behaviors similar to classic ferroelectrics such as the correlation between piezoelectricity and dielectricity \cite{LiFeiNatMat2018}. One of the most remarkable breakthroughs in relaxor ferroelectrics was the discovery of ultrahigh piezoelectricity which was designed by enhancing the dielectric response \cite{LiFeiAFM2018,LiFeiNatCommun2016}.  For perovskite ferroelectrics, the piezoelectric coefficient $d_{33}$ can be expressed as follows \cite{LiFeiAFM2017}
\begin{equation}
\label{eq:PiezoCoef}
d_{33} = 2Q_{33}P_r\varepsilon_{33}
\end{equation}
where $Q_{33}$ is the electrostrictive coefficient, $Pr$ the remanent polarization, and $\varepsilon_{33}$ the dielectric permittivity.  

For the perovskite relaxor ferroelectrics, the electrostrictive coefficients are similar ($Q_{33} \approx 0.8$ $m^4 C^{-2}$ ), so is the spontaneous polarization ($P_s \approx 0.4$ $C m^{-2}$ ) \cite{LiFeiAFM2017}. Hence, the piezoelectricity could be enhanced by improving the dielectric permittivity. However, the association of increase in $d_{33}$ with increase in $\varepsilon_{33}$ brought another issue that the ultra high dielectric response is affected by the nonlinear effect arising from the association of high piezoelectricity. Hence, it is of practical significance to  disarm the association of increase in $d_{33}$ with increase in $\varepsilon_{33}$. In the perovskite relaxor ferroelectrics, the enhanced dielectric response was achieved by flattening the thermodynamic energy profile between tetragonal ferroelectric domain and locally heterogeneous rhombohedral polar nanoregions (PNRs) \cite{LiFeiNatMat2018}.  The thermodynamic energy profile was flattened when the locally heterogeneous rhombohedral polar was able to achieve colinear with the polarization of tetragonal domain above the critical temperature denoted as $T_{colinear}$ that colinear PNRs begin to exist \cite{JIA2021JALCOM}. Thus, the locally heterogeneous rhombohedral PNRs could response to an external electric field or strain resulting in a enhanced dielectricity and piezoelectricity. However, if the temperature is sightly below $T_{colinear}$ , the locally heterogeneous rhombohedral polar would not aligned to polarization of tetragonal domain by the thermal energy but be able to be turned to consistent tropism with tetragonal polarization by external field, resulting in a mediocre piezoelectricity but improved dielectricity. Similar with the realization of enhanced piezoelectric responses using the nanoscale structural heterogeneity introduced through a doping strategy in PMN-PT \cite{LiFeiNatMat2018,LiFeiScience2019,JIA2021JALCOM}, disarming the association of increase in $d_{33}$ with increase in $\varepsilon_{33}$ was achieved by La doping in 0.4PNN-0.6PHT. 

In this work, La doped PNN-PHT samples were synthesized by solid phase method. Dielectric response depended on temperature was characterized to identify the critical temperature that locally heterogeneous rhombohedral polar was aligned with tetragonal domain. The micro domain morphology were observed through etching the surface of samples by diluted HF. In 3\% La doped $Pb(Ni_{1/3}Nb_{2/3})O_3-Pb(Hf_{0.3}Ti_{0.7})O_3$ (0.4PNN-0.6PHT), the dielectric permittivity increased 100\%, but the piezoelectric coefficient only increased 30\%.
\section{\label{sec:Exper}Experiment}
A series of  x at\% $La^{3+}$ doped 0.4PNN-0.6PHT (x=0, 0.25, 1, 2, 3, 4) were prepared by solid phase method \cite{LIU2018CerInt,YAN2018CerInt,YAN2020CerInt}. The precursor, $NiNb_2O_6$, was synthesized by sintering $NiO$ and $Nb_2O_5$  powders at 1100 $^\circ$C for 6h. $Pb_3O_4$(99.95\%, Aladdin), $TiO_2$ (99.99\%, Aladdin), $HfO_2$ (99.90\%, Macklin), $La_2O_3$(99.0\%, Aladdin) and $NiNb_2O_6$ were mixed evenly by ball-milling with zirconia balls for 12h. The mixed powders were calcined at 850 $^\circ$C for 2h, then, the secondary ball-milling was carried on for 12h. Using the grinding tool, the obtained powders were finally formed into a 10 mm diameter disc with a pressure of 200 MPa and sintered at 1250 $^\circ$C for two and a half hours. The final ceramic pellets were polished into 1mm thick discs with silver plated electrodes on both sides. For the piezoelectrical and dielectric properties, a 25 kV/cm external field was applied to the samples for 30 min in silicon oil. For observing the domain morphology, the cross section of the samples were etched in 5\% HF acid for few seconds.

The crystal structures of the samples were analyzed by X-ray diffraction (XRD, Rigaku, Japan). Besides, the micro domain morphology was observed using a scanning electron microscopy (SEM, JSM-6390A, JEOL Company). Dielectric properties were performed by an impedance analyzer (4287A, Agilent) from 25 $^\circ$C to 300 $^\circ$C . The hysteresis loop, electrostriction, and leakage current were obtained using a ferroelectric test system (aix ACCT, TF2000, Aachen, Germany) with a frequency of 1Hz at room temperature.  The piezoelectric constant was recorded by a quasi-static $d_{33}$ tester (YE2730A,Shijiao Technology Co., Ltd, China).
The domain structure and temperature dependent dielectric response were simulated via using phase-field model based on Ginzburg-Landau equation
\begin{equation}
\frac{\partial P_i(r,t)}{\partial t} = -L\frac{\delta F}{\delta P_i(r,t)}, (i = x, y, z)
\label{eq:TDGL}
\end{equation}
where $t$ denotes time, $L$ is the kinetic coefficient, $F$ is the total free energy of the system, $\frac{\delta F}{\delta P_i(r,t)}$ represents the thermodynamic driving force for the spatial and temporal evolution of the simulated system, and $r$ denotes the spatial vector, $r = (x, y , z)$.  This equation was solved by using a semi-implicit Fourier-spectral method considering the total free energy of the system included Landau energy term, elastic energy term, electrostatic energy term, and gradient energy term as described in the literatures \cite{CHEN1998CompuPhyCommun,Hu1998JACS,LiFeiNatCommun2016}.
\section{\label{sec:Results}Results and Discussions}
\begin{figure}
\includegraphics[width=0.8\linewidth]{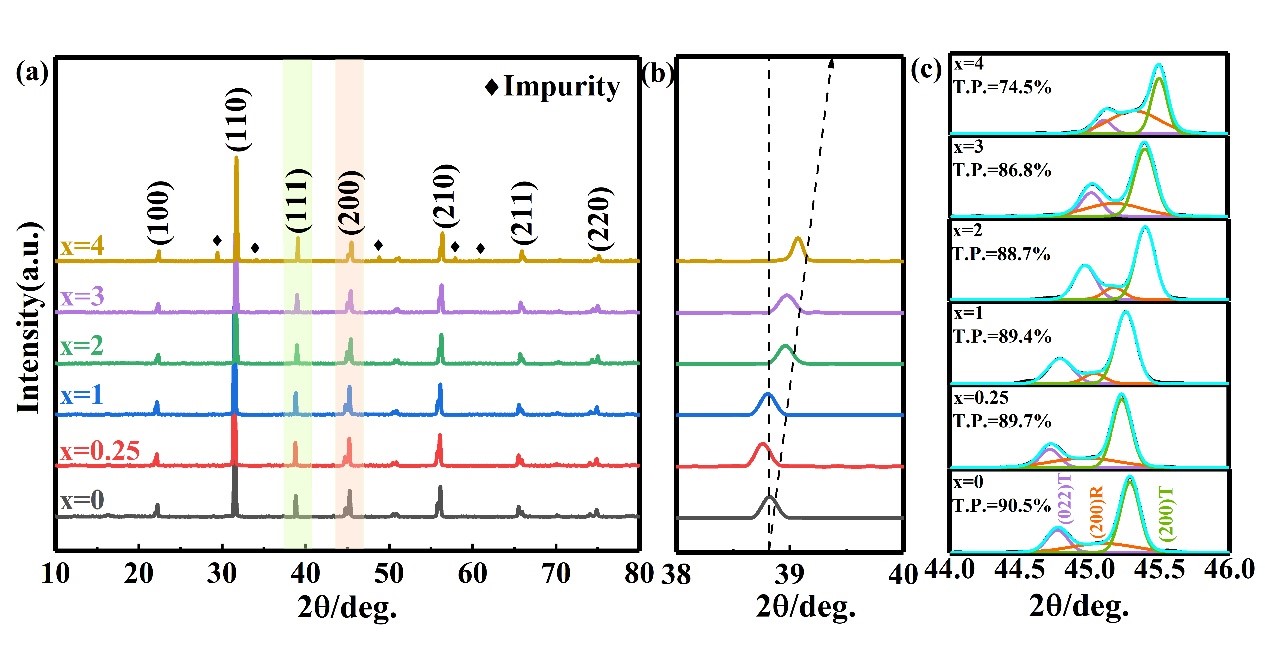}
\caption{\label{fig:XRD} (a) XRD patterns for 0.4PNN-0.6PHT with x at\% La doping (x= 0.25, 1, 2, 3 and 4), (b) locally amplified diffraction peaks near 39$^o$, (c) the fitting plots of peaks near 45$^o$.}
\end{figure}

Figure \ref{fig:XRD} describes the XRD diffraction patterns of 0.4PNN-0.6PHT with different $La$ doping concentration (0.25, 1, 2, 3, 4 at\%). Pure perovskite phase dominated in all as-prepared samples except when the concentration of $La$ is 4 at\%. La doped PNN-PHT shows a shift of (111) peak towards higher angles, as presented in Figure \ref{fig:XRD}b. The degree of deviation to higher diffraction angle deepens with the increasing amounts of $La$. This could be attributed to the fact that La ions are incorporated into PNN-PHT and replace Pb ions at A-sites of perovskite phase, and the diameter of $Pb^{2+}$ (1.19 \AA) \cite{ButeeJAD2016} is larger than that of $La^{3+}$ (1.061 \AA) \cite{TIANApplCata2009}. Further, in order to quantify the phase fraction in as-prepared La doped 0.4PNN-0.6PHT, the diffraction peaks between $44^o-46^o$ were obtained by slow scanning at the speed of $1^o$/min and were fitted using Gauss-Lorentz method. Figure \ref{fig:XRD}c shows the experimental and fitting results, which consisted of $(002)_R$ peak for R phase (orange curve), $(002)_T$ peak for T phase (green curve), and $(002)_T$ peak for T phase (purple curve). It reveals that the content of R phase increases with the increase of La content.

\begin{figure} 
\includegraphics[width=0.8\linewidth]{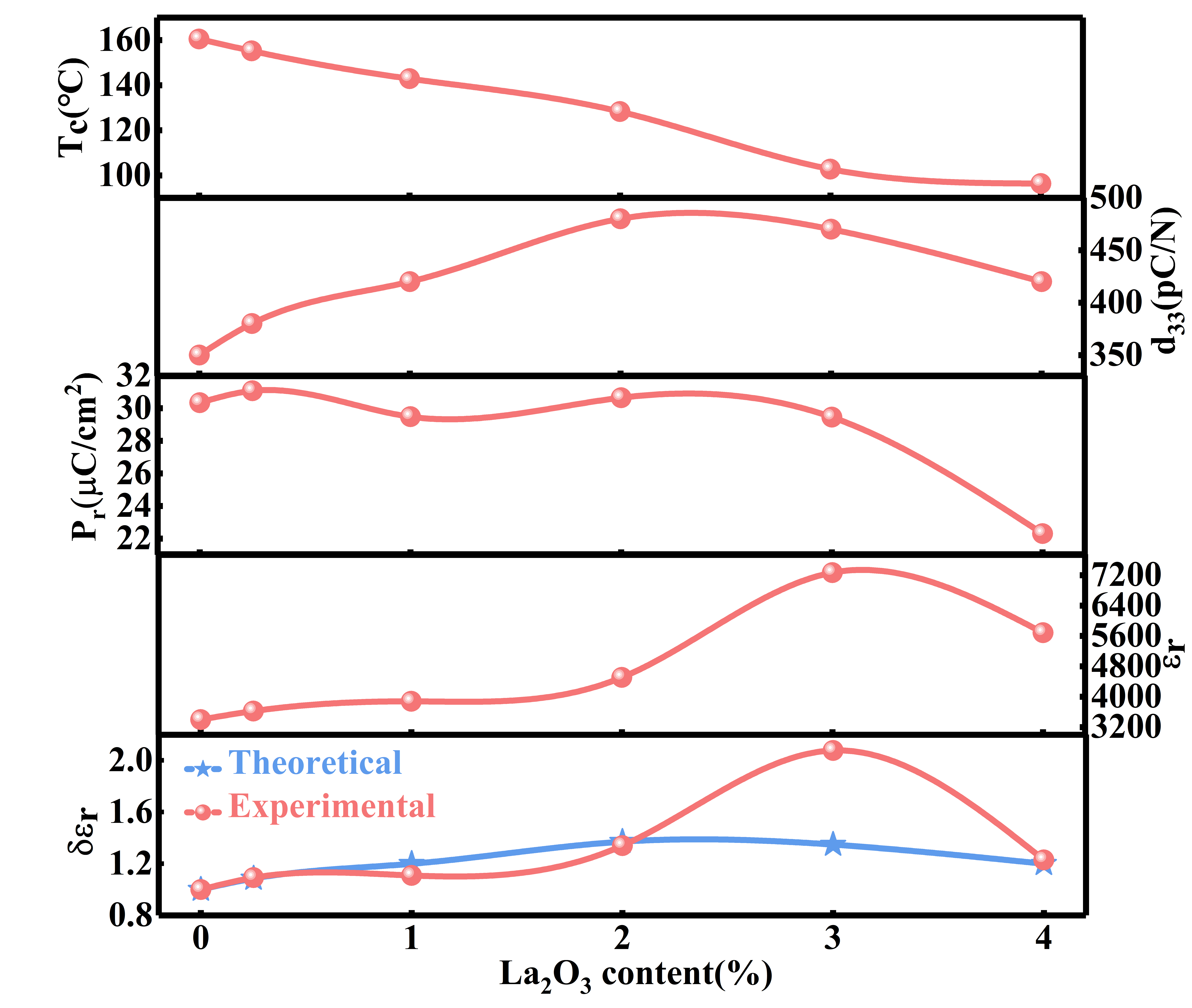}
\caption{\label{fig:Merge} Composition dependent of physical properties for La doped 0.4PNN-0.6PHT, (a)Curie temperature, $T_c$, (b)piezoelectric constant, $d_{33}$, (c)remanent polarization, $P_r$, (d)dielectric constant, $\varepsilon_{33}$, (e)comparison of $\varepsilon_{33}$ between theoretical and experimental results.}
\end{figure}

Figure \ref{fig:Merge} shows the behavior of 0.4PNN-0.6PHT depending on La doping concentration. Non significant synergies were observed among the profiles of Curie temperature $T_c$, piezoeletricity $d_{33}$, remanent polarization $P_r$ and dielectricity $\varepsilon_r$. As La doping concentration increased, the Curie temperature, gradually decreased from 160 $^\circ$C to 100 $^\circ$C, which were determined via dielectric temperature spectrum as shown in Supplementary Figure 1. The remanent polarization remained approximately the same value of 30 $\mu C/cm^2$ except for the 4\% La doped sample due to the existence of impurity phase, and the features of hysteresis loops for La doped samples are approximately the same as shown in Supplementary Figure 2. Both piezoelectricity and dielectricity increased first and then reduced with increase of La doping concentration. The max value of piezoelectricity was obtained at 2\% La doped, but dielectricity reached maximum at 3\% La doped. And most importantly, the increase in dielectricity and piezoelectricity differed and didn't follow Eq. \ref{eq:PiezoCoef} prediction when 3\% La was doped, even if impurity phase exists when 4\% La was doped, the association of increase in $d_{33}$ with increase in $\varepsilon_{33}$ was still hold. The disassociation of  increase in $d_{33}$ with increase in $\varepsilon_{33}$ for 3\% La doped  0.4PNN-0.6PHT was not caused by the variation of electrostrictive behaviors which did not demonstrate significant divergence as shown in Supplementary Figure 3 and 4. The increase in dielectricity of experimental measurement got significantly ahead of the theoretical prediction for 3\% La doped PNN-PHT. This abnormal improved dielectricity was attributed to the ferroelectric domain tailored by PNRs arising from La doping. When La doped, $La^{3+}$ would occupied at $Pb^{2+}$ site to form a local positive charge defect. Further, B-site vacancy would be generated to compensate $La_{Pb}^{\bullet}$ induced positive charge defect \cite{GuptaJAP1996},
\begin{equation}
\ce{xLa ->[{PNN-PHT}] xLa_{Pb}^{\cdot}+V_{B}^{''''}+ (4-x)h^{\cdot}}
\end{equation}

where B indicates the B-site cations in perovskite ABO$_3$. Hence, the charge compensation by B-site vacancy would cause hole carriers generation resulting in leak current reduction (Supplementary Figure 5), and B-site composition fluctuation resulting in enhanced dispersibility of phase transition (Supplementary Figure 1).

\begin{figure}
\includegraphics[width=1.0\linewidth]{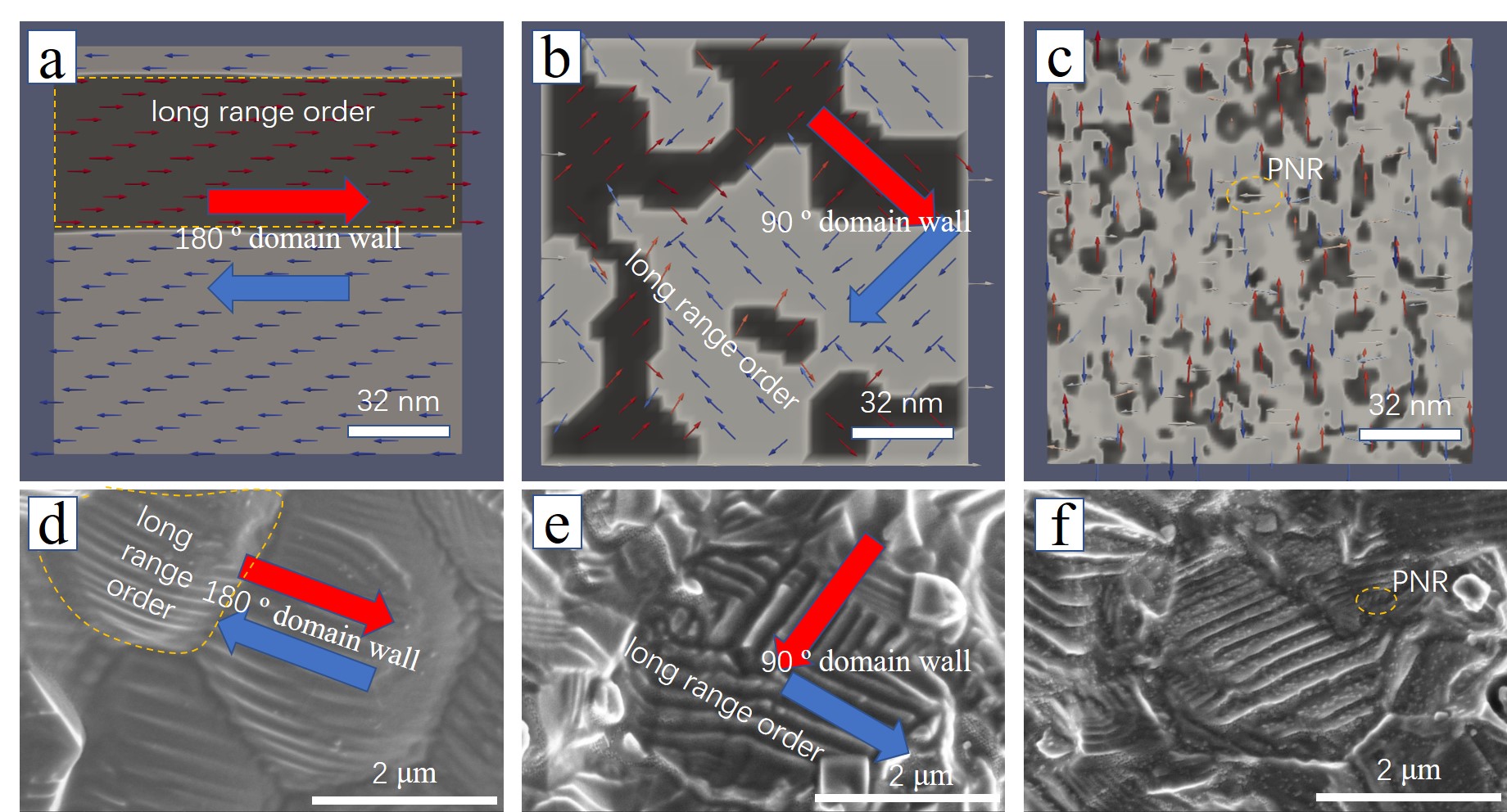}
\caption{\label{fig:DOMAIN} Domain morphology of 0.4PNN-0.6PHT with 0\%, 2\%, and 3\% La doping obtained by (a-c) simulation, and (d-e) SEM.}
\end{figure}

\begin{figure}
    \includegraphics[width = 1.0\linewidth]{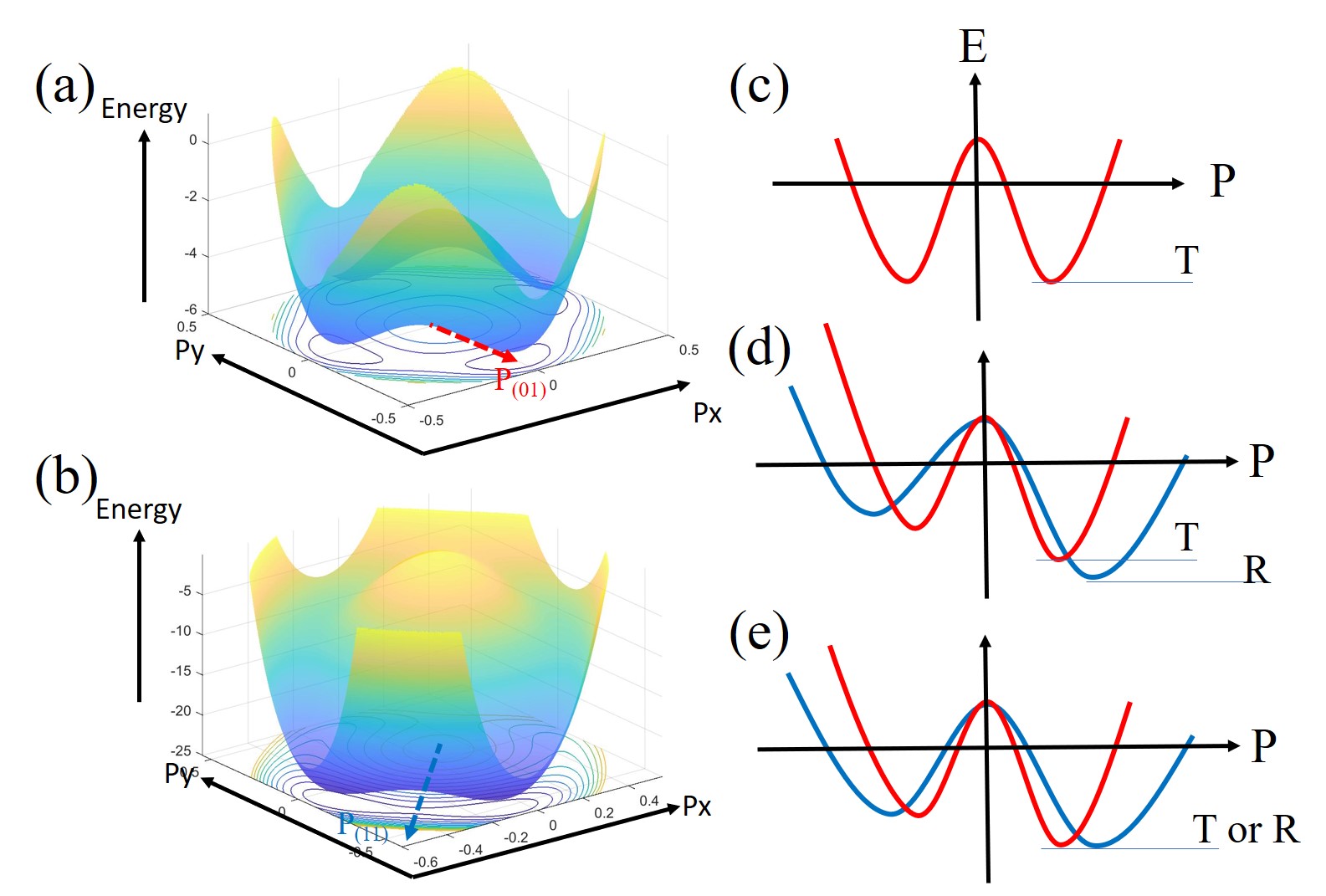}
    \caption{\label{fig:Energy}Schematic diagram of energy alignment in La doped PNN-PHT. (a) Energy Profile of PNN-PHT with tetragonal symmetry. (b) Energy profile of PNN-PHT with La doping. (c) Energy profile of PNN-PHT along the polarization. (d, e) Energy profile for tetragonal and rhombohedral phase before and after aligned.}
\end{figure}

\begin{figure}
\includegraphics[width=0.8\linewidth]{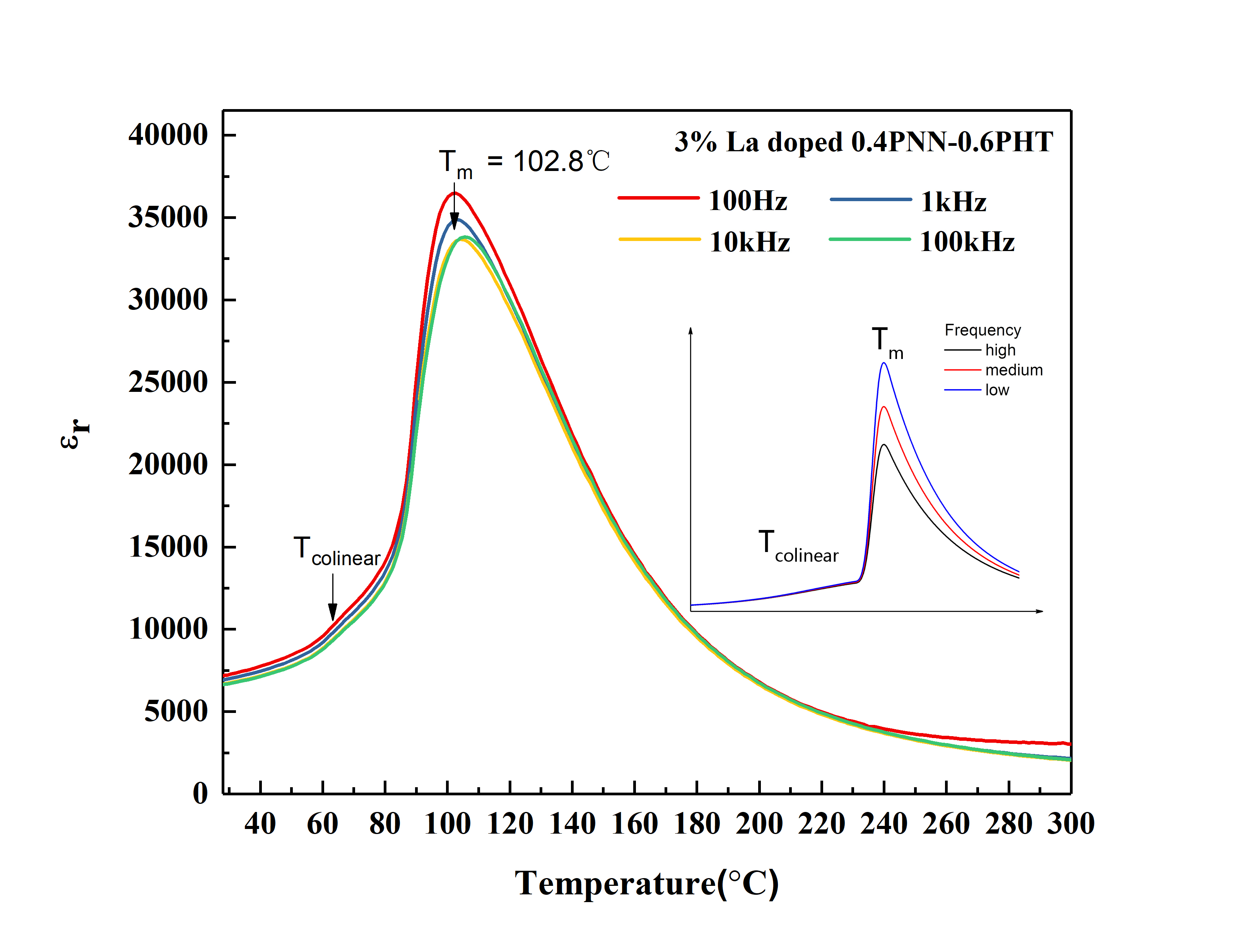}
\caption{\label{fig:EPSILON}Temperature dependence of dielectric permittivities for 3\% La doped. Inset indicates the simulated results.}
\end{figure}

According to research on Pb-based perovskite ferroelectrics, B-sites composition fluctuation would introduce effective random fields/bonds and/or change the ordering degree of B-site cation. Therefore, the addition of La into PNN–PHT is considered to be an effective approach to enhance its local structural heterogeneity, which is consider to be the origin of enhanced dielectricity. For PNN-PHT, it could be depicted as a tetragonal ferroelectric domain with locally introduced heterogeneous polar regions. In such a system, the bulk energy of those local polar regions, which is related to the local chemical composition, favours the rhombohedral phase (polar vectors along the diagonal $P_{(111)}$). The existence of local orthorhombic polar regions is equivalent to apply a field along (100) on the tetragonal ferroelectric domain. Hence, the local orthorhombic polar regions would change the morphology of the tetragonal ferroelectric domain. Figure \ref{fig:DOMAIN} shows the domain of PNN-PHT with different La doping concentration via theoretical simulation and  experimental observation. Without La doping, 0.4PNN-0.6PHT itself prefers tetragonal phase, which has been confirmed by XRD patterns as shown in Figure \ref{fig:XRD}. Due to the coupling of lattice and polarization in ferroelectric system, tetragonal ferroelectric was restricted forming a 180$^o$ domain wall while any perturbation didn't break the long range order as shown in Figure \ref{fig:DOMAIN}a and d. When a small amount of La was doped in 0.4PNN-0.6PHT, La occupied at Pb site and induced B-site vacancy. This charged defect pair constructed internal field along the diagonal $P_{(111)}$ and induced the appearance of rhombohedral phase. Due to the significant energy difference between tetragonal and rhombohedral phase, the long range order  was retained in tetragonal or rhombohedral phase, respectively. Thus, 90$^o$ domain wall was observed arsing from the existence of rhombohedral phase as shown in Figure \ref{fig:DOMAIN}b and e. When a moderate amount of La (3\%) was doped, the energy difference between tetragonal and rhombohedral phase was flattened. Long range order was broken, polar nano-regions (PNRs) appeared as shown in Figure \ref{fig:DOMAIN}c and f. Due to the easy collinearity of PNRs under applied field, an improved dielectricity was expected as shown in Figure \ref{fig:Merge}. 
To depict the properties of PNRs in the ‘collinear’ state, it is reasonable to use an electric field along [100] as the driving force to mimic the situation, where PNRs are allowed to rotate to the [100] direction to achieve collinear with tetragonal matrix. The Landau energy profile for tetragonal matrix and PNRs flattening and effect of collinear state of PNRs on dielectricity were presented in Figure \ref{fig:Energy} and \ref{fig:EPSILON}. Without La doping, 0.4PNN-0.6PHT energetically prefers tetragonal phase (Figure \ref{fig:Energy}a and c), while random field generated from PNN didn't break the long range order dominated by tetragonal phase. With La doped concentration below 2\%, effective random fields were improved. Thus,dispersibility through phase transition was enhanced (Supplementary Figure 2) by local structural heterogeneity arising from the B-site vacancies which formed defect dipolar along diagonal [111] orientation with La occupying at Pb sites. The defect dipolar along [111] would reduce the Landau energy of polarization along diagonal [111] orientation (Figure \ref{fig:Energy}d), inducing rhombohedral phase (Figure \ref{fig:Energy}b and Figure \ref{fig:XRD}). As La doping concentration increased to 3\%, the difference of Landau energy between tetragonal and rhombohedral phase was negligible (Figure \ref{fig:Energy}e), which could be conquered by energy of thermal noise. Thus, above a critical temperature, which was denoted as $T_{colinear}$, polarization could casually reorientate along either [100] or [111]. This arbitrariness of polarization broke the long range order, resulting in PNRs. Thus, enhanced dielectricity and piezoelectricity were expected. However, when the temperature was near below $T_{colinear}$, energy of thermal noise was not quite sufficient to conquer the energy barrier from rhombohedral phase to tetragonal phase. Long range order was partly preserved, and sporadic PNRs were observed (Figure \ref{fig:DOMAIN}f). Although energy of thermal noise was not quite sufficient to conquer the energy barrier to achieve the colinear of rhombohedral PNRs and tetragonal matrix, other energy could assist PNRs to stride over the energy barrier, for example electric field. Therefore, enhanced dielectricity but mediocre piezoelectricity was obtained (Figure and \ref{fig:Merge} and \ref{fig:EPSILON}).  

\section{\label{sec:Con}Conclusions}
0.4PNN-0.6PHT with different La doping concentration was synthesized. An abnormal improved dielectricity was observed where the association of increase in piezoelectricity with increase in dielectricity was destroyed. This abnormal improved dielectricity was investigated via micro-structure characterization by SEM and theoretical calculation by phase field method. It was attributed to the colinearity of rhombohedral PNRs among tetragonal matrix. This colinearity was arising from flattening energy difference between rhombohedral PNRs and tetragonal phase. This energy difference could be conquered by energy of thermal noise or external field. When temperature was near below $T_{colinear}$, enhanced dielectricity but mediocre piezoelectricity was obtained. 
\section*{Acknowledgement}
This work was supported by the National Natural Science Foundation of China (Grant No. 52173227, 61974114), and the Fundamental Research Funds for the Central Universities (Grand No. JB211404).

\nocite{*}

\bibliography{apssamp}

\providecommand{\noopsort}[1]{}\providecommand{\singleletter}[1]{#1}%
\begin{thebibliography}{23}%
\makeatletter
\providecommand \@ifxundefined [1]{%
 \@ifx{#1\undefined}
}%
\providecommand \@ifnum [1]{%
 \ifnum #1\expandafter \@firstoftwo
 \else \expandafter \@secondoftwo
 \fi
}%
\providecommand \@ifx [1]{%
 \ifx #1\expandafter \@firstoftwo
 \else \expandafter \@secondoftwo
 \fi
}%
\providecommand \natexlab [1]{#1}%
\providecommand \enquote  [1]{``#1''}%
\providecommand \bibnamefont  [1]{#1}%
\providecommand \bibfnamefont [1]{#1}%
\providecommand \citenamefont [1]{#1}%
\providecommand \href@noop [0]{\@secondoftwo}%
\providecommand \href [0]{\begingroup \@sanitize@url \@href}%
\providecommand \@href[1]{\@@startlink{#1}\@@href}%
\providecommand \@@href[1]{\endgroup#1\@@endlink}%
\providecommand \@sanitize@url [0]{\catcode `\\12\catcode `\$12\catcode
  `\&12\catcode `\#12\catcode `\^12\catcode `\_12\catcode `\%12\relax}%
\providecommand \@@startlink[1]{}%
\providecommand \@@endlink[0]{}%
\providecommand \url  [0]{\begingroup\@sanitize@url \@url }%
\providecommand \@url [1]{\endgroup\@href {#1}{\urlprefix }}%
\providecommand \urlprefix  [0]{URL }%
\providecommand \Eprint [0]{\href }%
\providecommand \doibase [0]{https://doi.org/}%
\providecommand \selectlanguage [0]{\@gobble}%
\providecommand \bibinfo  [0]{\@secondoftwo}%
\providecommand \bibfield  [0]{\@secondoftwo}%
\providecommand \translation [1]{[#1]}%
\providecommand \BibitemOpen [0]{}%
\providecommand \bibitemStop [0]{}%
\providecommand \bibitemNoStop [0]{.\EOS\space}%
\providecommand \EOS [0]{\spacefactor3000\relax}%
\providecommand \BibitemShut  [1]{\csname bibitem#1\endcsname}%
\let\auto@bib@innerbib\@empty
\bibitem [{\citenamefont {BOKOV}\ and\ \citenamefont
  {YE}(2012)}]{BOKOV2012JAD}%
  \BibitemOpen
  \bibfield  {author} {\bibinfo {author} {\bibfnamefont {A.~A.}\ \bibnamefont
  {BOKOV}}\ and\ \bibinfo {author} {\bibfnamefont {Z.-G.}\ \bibnamefont {YE}},\
  }\bibfield  {title} {\bibinfo {title} {Dielectric relaxation in relaxor
  ferroelectrics},\ }\href {https://doi.org/10.1142/S2010135X1241010X}
  {\bibfield  {journal} {\bibinfo  {journal} {Journal of Advanced Dielectrics}\
  }\textbf {\bibinfo {volume} {02}},\ \bibinfo {pages} {1241010} (\bibinfo
  {year} {2012})},\ \Eprint
  {https://arxiv.org/abs/https://doi.org/10.1142/S2010135X1241010X}
  {https://doi.org/10.1142/S2010135X1241010X} \BibitemShut {NoStop}%
\bibitem [{\citenamefont {Zhang}\ \emph
  {et~al.}(2021{\natexlab{a}})\citenamefont {Zhang}, \citenamefont {Xu},
  \citenamefont {Yue}, \citenamefont {Palma}, \citenamefont {Reece},\ and\
  \citenamefont {Yan}}]{ZHANG2021109447}%
  \BibitemOpen
  \bibfield  {author} {\bibinfo {author} {\bibfnamefont {M.}~\bibnamefont
  {Zhang}}, \bibinfo {author} {\bibfnamefont {X.}~\bibnamefont {Xu}}, \bibinfo
  {author} {\bibfnamefont {Y.}~\bibnamefont {Yue}}, \bibinfo {author}
  {\bibfnamefont {M.}~\bibnamefont {Palma}}, \bibinfo {author} {\bibfnamefont
  {M.~J.}\ \bibnamefont {Reece}},\ and\ \bibinfo {author} {\bibfnamefont
  {H.}~\bibnamefont {Yan}},\ }\bibfield  {title} {\bibinfo {title} {Multi
  elements substituted aurivillius phase relaxor ferroelectrics using high
  entropy design concept},\ }\href
  {https://doi.org/https://doi.org/10.1016/j.matdes.2020.109447} {\bibfield
  {journal} {\bibinfo  {journal} {Materials \& Design}\ }\textbf {\bibinfo
  {volume} {200}},\ \bibinfo {pages} {109447} (\bibinfo {year}
  {2021}{\natexlab{a}})}\BibitemShut {NoStop}%
\bibitem [{\citenamefont {Zhang}\ \emph
  {et~al.}(2021{\natexlab{b}})\citenamefont {Zhang}, \citenamefont {Liu},
  \citenamefont {Bokov}, \citenamefont {Zhang}, \citenamefont {Wang},
  \citenamefont {Ye},\ and\ \citenamefont {Jia}}]{ZhangPRB2021}%
  \BibitemOpen
  \bibfield  {author} {\bibinfo {author} {\bibfnamefont {J.}~\bibnamefont
  {Zhang}}, \bibinfo {author} {\bibfnamefont {L.}~\bibnamefont {Liu}}, \bibinfo
  {author} {\bibfnamefont {A.~A.}\ \bibnamefont {Bokov}}, \bibinfo {author}
  {\bibfnamefont {N.}~\bibnamefont {Zhang}}, \bibinfo {author} {\bibfnamefont
  {D.}~\bibnamefont {Wang}}, \bibinfo {author} {\bibfnamefont {Z.-G.}\
  \bibnamefont {Ye}},\ and\ \bibinfo {author} {\bibfnamefont {C.-L.}\
  \bibnamefont {Jia}},\ }\bibfield  {title} {\bibinfo {title} {Compositional
  ordering in relaxor ferroelectric
  $\mathrm{Pb}(b{B}^{\ensuremath{'}}){\mathrm{o}}_{3}$: Nearest neighbor
  approach},\ }\href {https://doi.org/10.1103/PhysRevB.103.054201} {\bibfield
  {journal} {\bibinfo  {journal} {Phys. Rev. B}\ }\textbf {\bibinfo {volume}
  {103}},\ \bibinfo {pages} {054201} (\bibinfo {year}
  {2021}{\natexlab{b}})}\BibitemShut {NoStop}%
\bibitem [{\citenamefont {Li}\ \emph {et~al.}(2018{\natexlab{a}})\citenamefont
  {Li}, \citenamefont {Lin}, \citenamefont {Chen}, \citenamefont {Cheng},
  \citenamefont {Wang}, \citenamefont {Li}, \citenamefont {Xu}, \citenamefont
  {Huang}, \citenamefont {Liao}, \citenamefont {Chen}, \citenamefont {Shrout},\
  and\ \citenamefont {Zhang}}]{LiFeiNatMat2018}%
  \BibitemOpen
  \bibfield  {author} {\bibinfo {author} {\bibfnamefont {F.}~\bibnamefont
  {Li}}, \bibinfo {author} {\bibfnamefont {D.}~\bibnamefont {Lin}}, \bibinfo
  {author} {\bibfnamefont {Z.}~\bibnamefont {Chen}}, \bibinfo {author}
  {\bibfnamefont {Z.}~\bibnamefont {Cheng}}, \bibinfo {author} {\bibfnamefont
  {J.}~\bibnamefont {Wang}}, \bibinfo {author} {\bibfnamefont {C.}~\bibnamefont
  {Li}}, \bibinfo {author} {\bibfnamefont {Z.}~\bibnamefont {Xu}}, \bibinfo
  {author} {\bibfnamefont {Q.}~\bibnamefont {Huang}}, \bibinfo {author}
  {\bibfnamefont {X.}~\bibnamefont {Liao}}, \bibinfo {author} {\bibfnamefont
  {L.-Q.}\ \bibnamefont {Chen}}, \bibinfo {author} {\bibfnamefont {T.~R.}\
  \bibnamefont {Shrout}},\ and\ \bibinfo {author} {\bibfnamefont
  {S.}~\bibnamefont {Zhang}},\ }\bibfield  {title} {\bibinfo {title} {Ultrahigh
  piezoelectricity in ferroelectric ceramics by design},\ }\href
  {https://doi.org/10.1038/s41563-018-0034-4} {\bibfield  {journal} {\bibinfo
  {journal} {Nature Mater.}\ }\textbf {\bibinfo {volume} {17}},\ \bibinfo
  {pages} {349–354} (\bibinfo {year} {2018}{\natexlab{a}})}\BibitemShut
  {NoStop}%
\bibitem [{\citenamefont {Li}\ \emph {et~al.}(2019)\citenamefont {Li},
  \citenamefont {Cabral}, \citenamefont {Xu}, \citenamefont {Cheng},
  \citenamefont {Dickey}, \citenamefont {LeBeau}, \citenamefont {Wang},
  \citenamefont {Luo}, \citenamefont {Taylor}, \citenamefont {Hackenberger},
  \citenamefont {Bellaiche}, \citenamefont {Xu}, \citenamefont {Chen},
  \citenamefont {Shrout},\ and\ \citenamefont {Zhang}}]{LiFeiScience2019}%
  \BibitemOpen
  \bibfield  {author} {\bibinfo {author} {\bibfnamefont {F.}~\bibnamefont
  {Li}}, \bibinfo {author} {\bibfnamefont {M.~J.}\ \bibnamefont {Cabral}},
  \bibinfo {author} {\bibfnamefont {B.}~\bibnamefont {Xu}}, \bibinfo {author}
  {\bibfnamefont {Z.}~\bibnamefont {Cheng}}, \bibinfo {author} {\bibfnamefont
  {E.~C.}\ \bibnamefont {Dickey}}, \bibinfo {author} {\bibfnamefont {J.~M.}\
  \bibnamefont {LeBeau}}, \bibinfo {author} {\bibfnamefont {J.}~\bibnamefont
  {Wang}}, \bibinfo {author} {\bibfnamefont {J.}~\bibnamefont {Luo}}, \bibinfo
  {author} {\bibfnamefont {S.}~\bibnamefont {Taylor}}, \bibinfo {author}
  {\bibfnamefont {W.}~\bibnamefont {Hackenberger}}, \bibinfo {author}
  {\bibfnamefont {L.}~\bibnamefont {Bellaiche}}, \bibinfo {author}
  {\bibfnamefont {Z.}~\bibnamefont {Xu}}, \bibinfo {author} {\bibfnamefont
  {L.-Q.}\ \bibnamefont {Chen}}, \bibinfo {author} {\bibfnamefont {T.~R.}\
  \bibnamefont {Shrout}},\ and\ \bibinfo {author} {\bibfnamefont
  {S.}~\bibnamefont {Zhang}},\ }\bibfield  {title} {\bibinfo {title} {Giant
  piezoelectricity of sm-doped $pb(mg_{1/3}nb_{2/3})o_3-pbtio_3$ single
  crystals},\ }\href {https://doi.org/10.1126/science.aaw2781} {\bibfield
  {journal} {\bibinfo  {journal} {Science}\ }\textbf {\bibinfo {volume}
  {364}},\ \bibinfo {pages} {264} (\bibinfo {year} {2019})},\ \Eprint
  {https://arxiv.org/abs/https://www.science.org/doi/pdf/10.1126/science.aaw2781}
  {https://www.science.org/doi/pdf/10.1126/science.aaw2781} \BibitemShut
  {NoStop}%
\bibitem [{\citenamefont {Yan}\ \emph {et~al.}(2021)\citenamefont {Yan},
  \citenamefont {Li}, \citenamefont {Jin}, \citenamefont {Du}, \citenamefont
  {Zhang}, \citenamefont {Zhang},\ and\ \citenamefont
  {Hao}}]{YangxiACSInterface2021}%
  \BibitemOpen
  \bibfield  {author} {\bibinfo {author} {\bibfnamefont {Y.}~\bibnamefont
  {Yan}}, \bibinfo {author} {\bibfnamefont {Z.}~\bibnamefont {Li}}, \bibinfo
  {author} {\bibfnamefont {L.}~\bibnamefont {Jin}}, \bibinfo {author}
  {\bibfnamefont {H.}~\bibnamefont {Du}}, \bibinfo {author} {\bibfnamefont
  {M.}~\bibnamefont {Zhang}}, \bibinfo {author} {\bibfnamefont
  {D.}~\bibnamefont {Zhang}},\ and\ \bibinfo {author} {\bibfnamefont
  {Y.}~\bibnamefont {Hao}},\ }\bibfield  {title} {\bibinfo {title} {Extremely
  high piezoelectric properties in pb-based ceramics through integrating phase
  boundary and defect engineering},\ }\href
  {https://doi.org/10.1021/acsami.1c10298} {\bibfield  {journal} {\bibinfo
  {journal} {ACS Applied Materials \& Interfaces}\ }\textbf {\bibinfo {volume}
  {13}},\ \bibinfo {pages} {38517} (\bibinfo {year} {2021})}\BibitemShut
  {NoStop}%
\bibitem [{\citenamefont {Chen}\ \emph {et~al.}(2019)\citenamefont {Chen},
  \citenamefont {Liu}, \citenamefont {Luo}, \citenamefont {Mei}, \citenamefont
  {Chen}, \citenamefont {Pan}, \citenamefont {Qi},\ and\ \citenamefont
  {Cao}}]{ChenJAP2019}%
  \BibitemOpen
  \bibfield  {author} {\bibinfo {author} {\bibfnamefont {Y.}~\bibnamefont
  {Chen}}, \bibinfo {author} {\bibfnamefont {K.~H.}\ \bibnamefont {Liu}},
  \bibinfo {author} {\bibfnamefont {Q.}~\bibnamefont {Luo}}, \bibinfo {author}
  {\bibfnamefont {M.}~\bibnamefont {Mei}}, \bibinfo {author} {\bibfnamefont
  {G.~L.}\ \bibnamefont {Chen}}, \bibinfo {author} {\bibfnamefont {R.~K.}\
  \bibnamefont {Pan}}, \bibinfo {author} {\bibfnamefont {Y.~J.}\ \bibnamefont
  {Qi}},\ and\ \bibinfo {author} {\bibfnamefont {W.~Q.}\ \bibnamefont {Cao}},\
  }\bibfield  {title} {\bibinfo {title} {Correlation of dielectric dispersion
  with distributed curie temperature in relaxor ferroelectrics},\ }\href
  {https://doi.org/10.1063/1.5080988} {\bibfield  {journal} {\bibinfo
  {journal} {Journal of Applied Physics}\ }\textbf {\bibinfo {volume} {125}},\
  \bibinfo {pages} {184104} (\bibinfo {year} {2019})}\BibitemShut {NoStop}%
\bibitem [{\citenamefont {Bobic}\ \emph {et~al.}(2018)\citenamefont {Bobic},
  \citenamefont {{Vijatovic Petrovic}},\ and\ \citenamefont
  {Stojanovic}}]{BOBIC2018233}%
  \BibitemOpen
  \bibfield  {author} {\bibinfo {author} {\bibfnamefont {J.~D.}\ \bibnamefont
  {Bobic}}, \bibinfo {author} {\bibfnamefont {M.~M.}\ \bibnamefont {{Vijatovic
  Petrovic}}},\ and\ \bibinfo {author} {\bibfnamefont {B.~D.}\ \bibnamefont
  {Stojanovic}},\ }\bibfield  {title} {\bibinfo {title} {11 - review of the
  most common relaxor ferroelectrics and their applications},\ }in\ \href
  {https://doi.org/https://doi.org/10.1016/B978-0-12-811180-2.00011-6} {\emph
  {\bibinfo {booktitle} {Magnetic, Ferroelectric, and Multiferroic Metal
  Oxides}}},\ \bibinfo {series and number} {Metal Oxides},\ \bibinfo {editor}
  {edited by\ \bibinfo {editor} {\bibfnamefont {B.~D.}\ \bibnamefont
  {Stojanovic}}}\ (\bibinfo  {publisher} {Elsevier},\ \bibinfo {year} {2018})\
  pp.\ \bibinfo {pages} {233--249}\BibitemShut {NoStop}%
\bibitem [{\citenamefont {Zhang}\ and\ \citenamefont
  {Huang}(2020)}]{ZhangSciRep2020}%
  \BibitemOpen
  \bibfield  {author} {\bibinfo {author} {\bibfnamefont {L.-L.}\ \bibnamefont
  {Zhang}}\ and\ \bibinfo {author} {\bibfnamefont {Y.-N.}\ \bibnamefont
  {Huang}},\ }\bibfield  {title} {\bibinfo {title} {Theory of
  relaxor-ferroelectricity},\ }\href
  {https://doi.org/10.1038/s41598-020-61911-5} {\bibfield  {journal} {\bibinfo
  {journal} {Sci. Rep.}\ }\textbf {\bibinfo {volume} {10}},\ \bibinfo {pages}
  {5060} (\bibinfo {year} {2020})}\BibitemShut {NoStop}%
\bibitem [{\citenamefont {Eremenko}\ \emph {et~al.}(2019)\citenamefont
  {Eremenko}, \citenamefont {Krayzman}, \citenamefont {Bosak}, \citenamefont
  {Playford}, \citenamefont {Chapman}, \citenamefont {Woicik}, \citenamefont
  {Ravel},\ and\ \citenamefont {Levin}}]{EremenkoNatCommun2019}%
  \BibitemOpen
  \bibfield  {author} {\bibinfo {author} {\bibfnamefont {M.}~\bibnamefont
  {Eremenko}}, \bibinfo {author} {\bibfnamefont {V.}~\bibnamefont {Krayzman}},
  \bibinfo {author} {\bibfnamefont {A.}~\bibnamefont {Bosak}}, \bibinfo
  {author} {\bibfnamefont {H.~Y.}\ \bibnamefont {Playford}}, \bibinfo {author}
  {\bibfnamefont {K.~W.}\ \bibnamefont {Chapman}}, \bibinfo {author}
  {\bibfnamefont {J.~C.}\ \bibnamefont {Woicik}}, \bibinfo {author}
  {\bibfnamefont {B.}~\bibnamefont {Ravel}},\ and\ \bibinfo {author}
  {\bibfnamefont {I.}~\bibnamefont {Levin}},\ }\bibfield  {title} {\bibinfo
  {title} {Local atomic order and hierarchical polar nanoregions in a classical
  relaxor ferroelectric},\ }\href {https://doi.org/10.1038/s41467-019-10665-4}
  {\bibfield  {journal} {\bibinfo  {journal} {Nat. Commun.}\ }\textbf {\bibinfo
  {volume} {10}},\ \bibinfo {pages} {2728} (\bibinfo {year}
  {2019})}\BibitemShut {NoStop}%
\bibitem [{\citenamefont {Li}\ \emph {et~al.}(2018{\natexlab{b}})\citenamefont
  {Li}, \citenamefont {Zhang}, \citenamefont {Damjanovic}, \citenamefont
  {Chen},\ and\ \citenamefont {Shrout}}]{LiFeiAFM2018}%
  \BibitemOpen
  \bibfield  {author} {\bibinfo {author} {\bibfnamefont {F.}~\bibnamefont
  {Li}}, \bibinfo {author} {\bibfnamefont {S.}~\bibnamefont {Zhang}}, \bibinfo
  {author} {\bibfnamefont {D.}~\bibnamefont {Damjanovic}}, \bibinfo {author}
  {\bibfnamefont {L.-Q.}\ \bibnamefont {Chen}},\ and\ \bibinfo {author}
  {\bibfnamefont {T.~R.}\ \bibnamefont {Shrout}},\ }\bibfield  {title}
  {\bibinfo {title} {Local structural heterogeneity and electromechanical
  responses of ferroelectrics: Learning from relaxor ferroelectrics},\
  }\href@noop {} {\bibfield  {journal} {\bibinfo  {journal} {Advanced
  Functional Materials}\ }\textbf {\bibinfo {volume} {28}},\ \bibinfo {pages}
  {1801504} (\bibinfo {year} {2018}{\natexlab{b}})}\BibitemShut {NoStop}%
\bibitem [{\citenamefont {Li}\ \emph {et~al.}(2016)\citenamefont {Li},
  \citenamefont {Zhang}, \citenamefont {Yang}, \citenamefont {Xu},
  \citenamefont {Zhang}, \citenamefont {Liu}, \citenamefont {Wang},
  \citenamefont {Wang}, \citenamefont {Cheng}, \citenamefont {Ye},
  \citenamefont {Luo}, \citenamefont {Shrout},\ and\ \citenamefont
  {Chen}}]{LiFeiNatCommun2016}%
  \BibitemOpen
  \bibfield  {author} {\bibinfo {author} {\bibfnamefont {F.}~\bibnamefont
  {Li}}, \bibinfo {author} {\bibfnamefont {S.}~\bibnamefont {Zhang}}, \bibinfo
  {author} {\bibfnamefont {T.}~\bibnamefont {Yang}}, \bibinfo {author}
  {\bibfnamefont {Z.}~\bibnamefont {Xu}}, \bibinfo {author} {\bibfnamefont
  {N.}~\bibnamefont {Zhang}}, \bibinfo {author} {\bibfnamefont
  {G.}~\bibnamefont {Liu}}, \bibinfo {author} {\bibfnamefont {J.}~\bibnamefont
  {Wang}}, \bibinfo {author} {\bibfnamefont {J.}~\bibnamefont {Wang}}, \bibinfo
  {author} {\bibfnamefont {Z.}~\bibnamefont {Cheng}}, \bibinfo {author}
  {\bibfnamefont {Z.-G.}\ \bibnamefont {Ye}}, \bibinfo {author} {\bibfnamefont
  {J.}~\bibnamefont {Luo}}, \bibinfo {author} {\bibfnamefont {T.~R.}\
  \bibnamefont {Shrout}},\ and\ \bibinfo {author} {\bibfnamefont {L.-Q.}\
  \bibnamefont {Chen}},\ }\bibfield  {title} {\bibinfo {title} {The origin of
  ultrahigh piezoelectricity in relaxor-ferroelectric solid solution
  crystals},\ }\href {https://doi.org/10.1038/ncomms13807} {\bibfield
  {journal} {\bibinfo  {journal} {Nat. Commun.}\ }\textbf {\bibinfo {volume}
  {7}},\ \bibinfo {pages} {13807} (\bibinfo {year} {2016})}\BibitemShut
  {NoStop}%
\bibitem [{\citenamefont {Li}\ \emph {et~al.}(2017)\citenamefont {Li},
  \citenamefont {Zhang}, \citenamefont {Xu},\ and\ \citenamefont
  {Chen}}]{LiFeiAFM2017}%
  \BibitemOpen
  \bibfield  {author} {\bibinfo {author} {\bibfnamefont {F.}~\bibnamefont
  {Li}}, \bibinfo {author} {\bibfnamefont {S.}~\bibnamefont {Zhang}}, \bibinfo
  {author} {\bibfnamefont {Z.}~\bibnamefont {Xu}},\ and\ \bibinfo {author}
  {\bibfnamefont {L.-Q.}\ \bibnamefont {Chen}},\ }\bibfield  {title} {\bibinfo
  {title} {The contributions of polar nanoregions to the dielectric and
  piezoelectric responses in domain-engineered relaxor-pbtio3 crystals},\
  }\href@noop {} {\bibfield  {journal} {\bibinfo  {journal} {Advanced
  Functional Materials}\ }\textbf {\bibinfo {volume} {27}},\ \bibinfo {pages}
  {1700310} (\bibinfo {year} {2017})}\BibitemShut {NoStop}%
\bibitem [{\citenamefont {Jia}\ \emph {et~al.}(2021)\citenamefont {Jia},
  \citenamefont {Yang}, \citenamefont {Zhu}, \citenamefont {Li},\ and\
  \citenamefont {Wang}}]{JIA2021JALCOM}%
  \BibitemOpen
  \bibfield  {author} {\bibinfo {author} {\bibfnamefont {H.}~\bibnamefont
  {Jia}}, \bibinfo {author} {\bibfnamefont {S.}~\bibnamefont {Yang}}, \bibinfo
  {author} {\bibfnamefont {W.}~\bibnamefont {Zhu}}, \bibinfo {author}
  {\bibfnamefont {F.}~\bibnamefont {Li}},\ and\ \bibinfo {author}
  {\bibfnamefont {L.}~\bibnamefont {Wang}},\ }\bibfield  {title} {\bibinfo
  {title} {Improved piezoelectric properties of pb(mg1/3nb2/3)o3-pbtio3
  textured ferroelectric ceramics via sm-doping method},\ }\href
  {https://doi.org/https://doi.org/10.1016/j.jallcom.2021.160666} {\bibfield
  {journal} {\bibinfo  {journal} {Journal of Alloys and Compounds}\ }\textbf
  {\bibinfo {volume} {881}},\ \bibinfo {pages} {160666} (\bibinfo {year}
  {2021})}\BibitemShut {NoStop}%
\bibitem [{\citenamefont {Liu}\ \emph {et~al.}(2018)\citenamefont {Liu},
  \citenamefont {Li}, \citenamefont {Yan}, \citenamefont {Zhang}, \citenamefont
  {Zhang},\ and\ \citenamefont {Hao}}]{LIU2018CerInt}%
  \BibitemOpen
  \bibfield  {author} {\bibinfo {author} {\bibfnamefont {Z.}~\bibnamefont
  {Liu}}, \bibinfo {author} {\bibfnamefont {Z.}~\bibnamefont {Li}}, \bibinfo
  {author} {\bibfnamefont {Y.}~\bibnamefont {Yan}}, \bibinfo {author}
  {\bibfnamefont {M.}~\bibnamefont {Zhang}}, \bibinfo {author} {\bibfnamefont
  {D.}~\bibnamefont {Zhang}},\ and\ \bibinfo {author} {\bibfnamefont
  {Y.}~\bibnamefont {Hao}},\ }\bibfield  {title} {\bibinfo {title} {Low
  temperature sintering of pb(ni, nb)o3-pb(hf, ti)o3 piezoceramics with fe2o3
  and bi2o3 addition},\ }\href
  {https://doi.org/https://doi.org/10.1016/j.ceramint.2018.08.321} {\bibfield
  {journal} {\bibinfo  {journal} {Ceramics International}\ }\textbf {\bibinfo
  {volume} {44}},\ \bibinfo {pages} {23263} (\bibinfo {year}
  {2018})}\BibitemShut {NoStop}%
\bibitem [{\citenamefont {Yan}\ \emph {et~al.}(2018)\citenamefont {Yan},
  \citenamefont {Liu}, \citenamefont {Li}, \citenamefont {Zhang}, \citenamefont
  {Zhang},\ and\ \citenamefont {Feng}}]{YAN2018CerInt}%
  \BibitemOpen
  \bibfield  {author} {\bibinfo {author} {\bibfnamefont {Y.}~\bibnamefont
  {Yan}}, \bibinfo {author} {\bibfnamefont {Z.}~\bibnamefont {Liu}}, \bibinfo
  {author} {\bibfnamefont {Z.}~\bibnamefont {Li}}, \bibinfo {author}
  {\bibfnamefont {M.}~\bibnamefont {Zhang}}, \bibinfo {author} {\bibfnamefont
  {D.}~\bibnamefont {Zhang}},\ and\ \bibinfo {author} {\bibfnamefont
  {Y.}~\bibnamefont {Feng}},\ }\bibfield  {title} {\bibinfo {title} {Improving
  piezoelectric properties of pb(ni, nb)o3-pb(hf, ti)o3 ceramics by lif
  addition},\ }\href
  {https://doi.org/https://doi.org/10.1016/j.ceramint.2017.12.094} {\bibfield
  {journal} {\bibinfo  {journal} {Ceramics International}\ }\textbf {\bibinfo
  {volume} {44}},\ \bibinfo {pages} {5790} (\bibinfo {year}
  {2018})}\BibitemShut {NoStop}%
\bibitem [{\citenamefont {Yan}\ \emph {et~al.}(2020)\citenamefont {Yan},
  \citenamefont {Li}, \citenamefont {Xia}, \citenamefont {Zhao}, \citenamefont
  {Zhang},\ and\ \citenamefont {Zhang}}]{YAN2020CerInt}%
  \BibitemOpen
  \bibfield  {author} {\bibinfo {author} {\bibfnamefont {Y.}~\bibnamefont
  {Yan}}, \bibinfo {author} {\bibfnamefont {Z.}~\bibnamefont {Li}}, \bibinfo
  {author} {\bibfnamefont {Y.}~\bibnamefont {Xia}}, \bibinfo {author}
  {\bibfnamefont {M.}~\bibnamefont {Zhao}}, \bibinfo {author} {\bibfnamefont
  {M.}~\bibnamefont {Zhang}},\ and\ \bibinfo {author} {\bibfnamefont
  {D.}~\bibnamefont {Zhang}},\ }\bibfield  {title} {\bibinfo {title}
  {Ultra-high piezoelectric and dielectric properties of
  low-temperature-sintered lead hafnium titanate-lead niobium nickelate
  ceramics},\ }\href
  {https://doi.org/https://doi.org/10.1016/j.ceramint.2019.10.166} {\bibfield
  {journal} {\bibinfo  {journal} {Ceramics International}\ }\textbf {\bibinfo
  {volume} {46}},\ \bibinfo {pages} {5448} (\bibinfo {year}
  {2020})}\BibitemShut {NoStop}%
\bibitem [{\citenamefont {Chen}\ and\ \citenamefont
  {Shen}(1998)}]{CHEN1998CompuPhyCommun}%
  \BibitemOpen
  \bibfield  {author} {\bibinfo {author} {\bibfnamefont {L.}~\bibnamefont
  {Chen}}\ and\ \bibinfo {author} {\bibfnamefont {J.}~\bibnamefont {Shen}},\
  }\bibfield  {title} {\bibinfo {title} {Applications of semi-implicit
  fourier-spectral method to phase field equations},\ }\href
  {https://doi.org/https://doi.org/10.1016/S0010-4655(97)00115-X} {\bibfield
  {journal} {\bibinfo  {journal} {Computer Physics Communications}\ }\textbf
  {\bibinfo {volume} {108}},\ \bibinfo {pages} {147} (\bibinfo {year}
  {1998})}\BibitemShut {NoStop}%
\bibitem [{\citenamefont {Hu}\ and\ \citenamefont {Chen}(1998)}]{Hu1998JACS}%
  \BibitemOpen
  \bibfield  {author} {\bibinfo {author} {\bibfnamefont {H.-L.}\ \bibnamefont
  {Hu}}\ and\ \bibinfo {author} {\bibfnamefont {L.-Q.}\ \bibnamefont {Chen}},\
  }\bibfield  {title} {\bibinfo {title} {Three-dimensional computer simulation
  of ferroelectric domain formation},\ }\href
  {https://doi.org/https://doi.org/10.1111/j.1151-2916.1998.tb02367.x}
  {\bibfield  {journal} {\bibinfo  {journal} {Journal of the American Ceramic
  Society}\ }\textbf {\bibinfo {volume} {81}},\ \bibinfo {pages} {492}
  (\bibinfo {year} {1998})},\ \Eprint
  {https://arxiv.org/abs/https://ceramics.onlinelibrary.wiley.com/doi/pdf/10.1111/j.1151-2916.1998.tb02367.x}
  {https://ceramics.onlinelibrary.wiley.com/doi/pdf/10.1111/j.1151-2916.1998.tb02367.x}
  \BibitemShut {NoStop}%
\bibitem [{\citenamefont {Butee}\ \emph {et~al.}(2016)\citenamefont {Butee},
  \citenamefont {Kambale}, \citenamefont {Upadhyay}, \citenamefont {Bashaiah},
  \citenamefont {Raju},\ and\ \citenamefont {Panda}}]{ButeeJAD2016}%
  \BibitemOpen
  \bibfield  {author} {\bibinfo {author} {\bibfnamefont {S.~P.}\ \bibnamefont
  {Butee}}, \bibinfo {author} {\bibfnamefont {K.~R.}\ \bibnamefont {Kambale}},
  \bibinfo {author} {\bibfnamefont {S.}~\bibnamefont {Upadhyay}}, \bibinfo
  {author} {\bibfnamefont {S.}~\bibnamefont {Bashaiah}}, \bibinfo {author}
  {\bibfnamefont {K.~C.~J.}\ \bibnamefont {Raju}},\ and\ \bibinfo {author}
  {\bibfnamefont {H.}~\bibnamefont {Panda}},\ }\bibfield  {title} {\bibinfo
  {title} {Synthesis and microwave dielectric behavior of
  $(bi_{1-x}xpb_x)nbo_4$ ceramics},\ }\href
  {https://doi.org/10.1142/S2010135X16500065} {\bibfield  {journal} {\bibinfo
  {journal} {Journal of Advanced Dielectrics}\ }\textbf {\bibinfo {volume}
  {06}},\ \bibinfo {pages} {1650006} (\bibinfo {year} {2016})}\BibitemShut
  {NoStop}%
\bibitem [{\citenamefont {Tian}\ \emph {et~al.}(2009)\citenamefont {Tian},
  \citenamefont {Wang}, \citenamefont {Wang}, \citenamefont {Zhu},\ and\
  \citenamefont {Zhang}}]{TIANApplCata2009}%
  \BibitemOpen
  \bibfield  {author} {\bibinfo {author} {\bibfnamefont {M.}~\bibnamefont
  {Tian}}, \bibinfo {author} {\bibfnamefont {A.}~\bibnamefont {Wang}}, \bibinfo
  {author} {\bibfnamefont {X.}~\bibnamefont {Wang}}, \bibinfo {author}
  {\bibfnamefont {Y.}~\bibnamefont {Zhu}},\ and\ \bibinfo {author}
  {\bibfnamefont {T.}~\bibnamefont {Zhang}},\ }\bibfield  {title} {\bibinfo
  {title} {Effect of large cations (la3+ and ba2+) on the catalytic performance
  of mn-substituted hexaaluminates for n2o decomposition},\ }\href
  {https://doi.org/https://doi.org/10.1016/j.apcatb.2009.09.002} {\bibfield
  {journal} {\bibinfo  {journal} {Applied Catalysis B: Environmental}\ }\textbf
  {\bibinfo {volume} {92}},\ \bibinfo {pages} {437} (\bibinfo {year}
  {2009})}\BibitemShut {NoStop}%
\bibitem [{\citenamefont {Gupta}\ and\ \citenamefont
  {Viehland}(1996)}]{GuptaJAP1996}%
  \BibitemOpen
  \bibfield  {author} {\bibinfo {author} {\bibfnamefont {S.~M.}\ \bibnamefont
  {Gupta}}\ and\ \bibinfo {author} {\bibfnamefont {D.}~\bibnamefont
  {Viehland}},\ }\bibfield  {title} {\bibinfo {title} {Role of charge
  compensation mechanism in la‐modified pb(mg1/3nb2/3)o3–pbtio3 ceramics:
  Enhanced ordering and pyrochlore formation},\ }\href
  {https://doi.org/10.1063/1.363581} {\bibfield  {journal} {\bibinfo  {journal}
  {Journal of Applied Physics}\ }\textbf {\bibinfo {volume} {80}},\ \bibinfo
  {pages} {5875} (\bibinfo {year} {1996})}\BibitemShut {NoStop}%
\bibitem [{\citenamefont {Sheng}\ \emph {et~al.}(2008)\citenamefont {Sheng},
  \citenamefont {Zhang}, \citenamefont {Li}, \citenamefont {Choudhury},
  \citenamefont {Jia}, \citenamefont {Liu},\ and\ \citenamefont
  {Chen}}]{Shen2008JAP}%
  \BibitemOpen
  \bibfield  {author} {\bibinfo {author} {\bibfnamefont {G.}~\bibnamefont
  {Sheng}}, \bibinfo {author} {\bibfnamefont {J.~X.}\ \bibnamefont {Zhang}},
  \bibinfo {author} {\bibfnamefont {Y.~L.}\ \bibnamefont {Li}}, \bibinfo
  {author} {\bibfnamefont {S.}~\bibnamefont {Choudhury}}, \bibinfo {author}
  {\bibfnamefont {Q.~X.}\ \bibnamefont {Jia}}, \bibinfo {author} {\bibfnamefont
  {Z.~K.}\ \bibnamefont {Liu}},\ and\ \bibinfo {author} {\bibfnamefont {L.~Q.}\
  \bibnamefont {Chen}},\ }\bibfield  {title} {\bibinfo {title} {Domain
  stability of pbtio3 thin films under anisotropic misfit strains: Phase-field
  simulations},\ }\href {https://doi.org/10.1063/1.2974093} {\bibfield
  {journal} {\bibinfo  {journal} {Journal of Applied Physics}\ }\textbf
  {\bibinfo {volume} {104}},\ \bibinfo {pages} {054105} (\bibinfo {year}
  {2008})}\BibitemShut {NoStop}%
\end{thebibliography}%

\end{document}


\begin{figure}[H]
    \centering
    \includegraphics[width=1.0\linewidth]{FigS/Figure S2_epsilon.jpg}
    \caption{Dielectric temperature spectrum for 0.4PNN-0.6PHT with and without La doping, inset shows the fitted dispersion factor $\gamma$.}
    \label{fig:epsilon}
\end{figure}
\begin{figure}[H]
    \centering
    \includegraphics[width=0.8\linewidth]{FigS/Figure S2_Hysteresis loop.jpg}
    \caption{Hysteresis loops for 0.4PNN-0.6PHT with and without La doping.}
    \label{fig:hysteresis}
\end{figure}

\begin{figure}[H]
    \centering
    \includegraphics[width=0.8\linewidth]{FigS/Figure S3_strain.jpg}
    \caption{ Bipolar S-E loops for 0.4PNN-0.6PHT with and without La doping. }
    \label{fig:strain}
\end{figure}
\begin{figure}[H]
    \centering
    \includegraphics[width=0.8\linewidth]{FigS/Figure S4_strain_single.jpg}
    \caption{Electrostriction curves for 0.4PNN-0.6PHT with and without La doping. }
    \label{fig:single}
\end{figure}
\begin{figure}[H]
    \centering
    \includegraphics[width=0.8\linewidth]{FigS/Figure S1_leak.jpg}
    \caption{Leak current for 0.4PNN-0.6PHT with and without La doping under 1 kV/cm and 2 kV/cm electric field.}
    \label{fig:leak}
\end{figure}